\documentclass{aa}
\usepackage{txfonts}
\usepackage{graphicx}
\date{Received <date> / Accepted <date>}
\begin{document}
\title{ First detections of extragalactic SO$_2$, NS and NO}
\subtitle{}
\author{S. Mart\'{\i}n \inst{1} \and R. Mauersberger \inst{1} \and J. Mart\'{\i}n-Pintado \inst{2} \and S. Garc\'{\i}a-Burillo \inst{3} \and C. Henkel \inst{4}}
\institute{Instituto de Radioastronom\'{\i}a Milim\'etrica (IRAM), Avda. Divina Pastora 7 NC, E-18012 Granada, Spain
		\and
	Departamento de Astrofis\'{\i}ca Molecular e Infrarroja, Instituto de Estructura de la Materia, CSIC, Serrano 121, E-28006 Madrid, Spain
		\and
	Observatorio Astron\'omico Nacional (OAN), Apartado 1143, 28800 Alcal\'a de Henares, Madrid, Spain
		\and
	Max-Planck-Institut f\"ur Radioastronomie, Auf dem H\"ugel 69, D-53121 Bonn, Germany
	}
\offprints{S. Mart\'{\i}n, \email{martin@iram.es}}
\date{Received <date> / Accepted <date>}
\abstract{We report the first detections of SO$_2$, NS and NO in an extragalactic source, the nucleus of the starburst galaxy \object{NGC\,253}.
Five SO$_2$ transitions, three groups of hyperfine components of NO and five of NS were detected.
All three species show large abundances averaged over the inner 200\,pc of \object{NGC\,253}.
With a relative abundance of a few $10^{-7}$, the emission of the NO molecule is similar or even larger than that found in Galactic
star forming regions.
The derived relative molecular abundances for each molecule have been compared with those of prototypical Galactic molecular clouds.
These results seem to confirm that large scale shocks dominate the chemistry of these molecules in the nucleus of \object{NGC\,253}, ruling out a
chemistry dominated by PDRs for the bulk of the gas.
\keywords{ISM: molecules - galaxies,individual: NGC\,253 -- galaxies: ISM -- galaxies: starburst -- galaxies: abundances} }%
\maketitle
\section{Introduction}
Molecular lines from the central regions of galaxies are among the most promising tools to explore and understand the
history of our universe (e.g. Combes et al. \cite{Combes} and references therein). In many galaxies most of the
molecular emission stems from highly excited gas within the central few 100 pc. However, the processes driving the excitation and the complex
chemistry of the gas are not entirely clear.

Molecular studies of nearby active galaxies such as \object{NGC\,253}, \object{M\,82} or \object{Arp\,220} suggest that C shocks, photodissociating radiation, and
irradiation by X rays or cosmic rays (e.g. Rigopoulou et al. \cite{Rigopoulou}, G\"usten et al. \cite{Gusten}, Farquhar et al. \cite{Farquhar}) play an important role
in the heating and the chemistry of nuclear gas.
The dominant mechanisms and their relative importance are still unclear.

Sulfur-bearing molecules such as SO$_2$ and NS are present in a wide  variety of interstellar conditions, displaying enhanced abundances in
the hot cores of high mass star forming regions.
In hot cores, the chemistry of these molecules is determined by grain-mantle evaporation into warm gas
(Charnley, \cite{Charnley}). Abundance ratios such as NS/CS have been suggested to be one of the signatures of shocks in hot cores (Viti et al. \cite{Viti}).
The NO molecule is a fairly abundant molecule in the diffuse ISM of the Milky Way.
This species is the main precursor of the N/O chemical network (Halfen et al. \cite{Halfen}).

In this Letter we report the first detections of SO$_2$, NS and NO in an extragalactic source, the nucleus of \object{NGC\,253}.
This region is one of the most prolific sources of molecular emission outside the Milky Way (Mauersberger \& Henkel \cite{Mauers93}).
The large amount of hot gas (Mauersberger et al. \cite{Mauers03}) and the high abundance of molecules with special chemistry, such as SiO
(Garc\'{\i}a-Burillo et al. \cite{Burillo}) indicates that the presence of large scale shocks should dominate the heating and the chemistry of the
molecular gas.
Also the very high abundance of the three molecules reported in this Letter
and their excitation can be explained if the chemistry
of the nucleus of \object{NGC\,253} is dominated by large scale shocks.

\section{Observations and results}
Observations of the spectral lines (Table~\ref{tab:fits}) were carried out
with the IRAM 30m telescope.
Continuum measurements on nearby sources made every $\sim\!2$\,hours were used to keep a pointing
accuracy better than $\sim\!3\arcsec$. A wobbling secondary mirror was used, with a
symmetrical beam throw of $4\arcmin$ in azimuth and a switching frequency of 0.5\,Hz.

\begin{figure}[!tb]
	\begin{minipage}[b]{0.23\textwidth}
		\includegraphics[width=\linewidth]{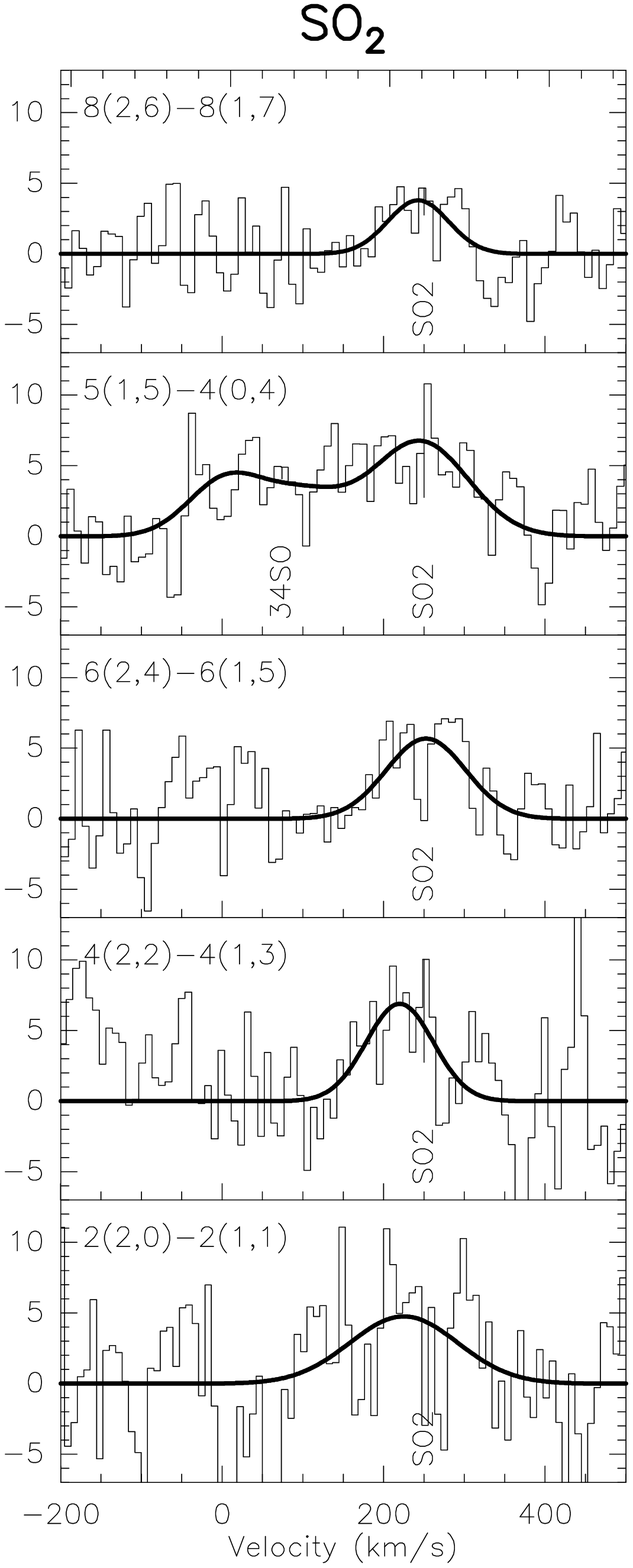}
	\end{minipage}%
	\hspace{5 pt}
	\begin{minipage}[b]{0.23\textwidth}
			\includegraphics[width=1.035\linewidth]{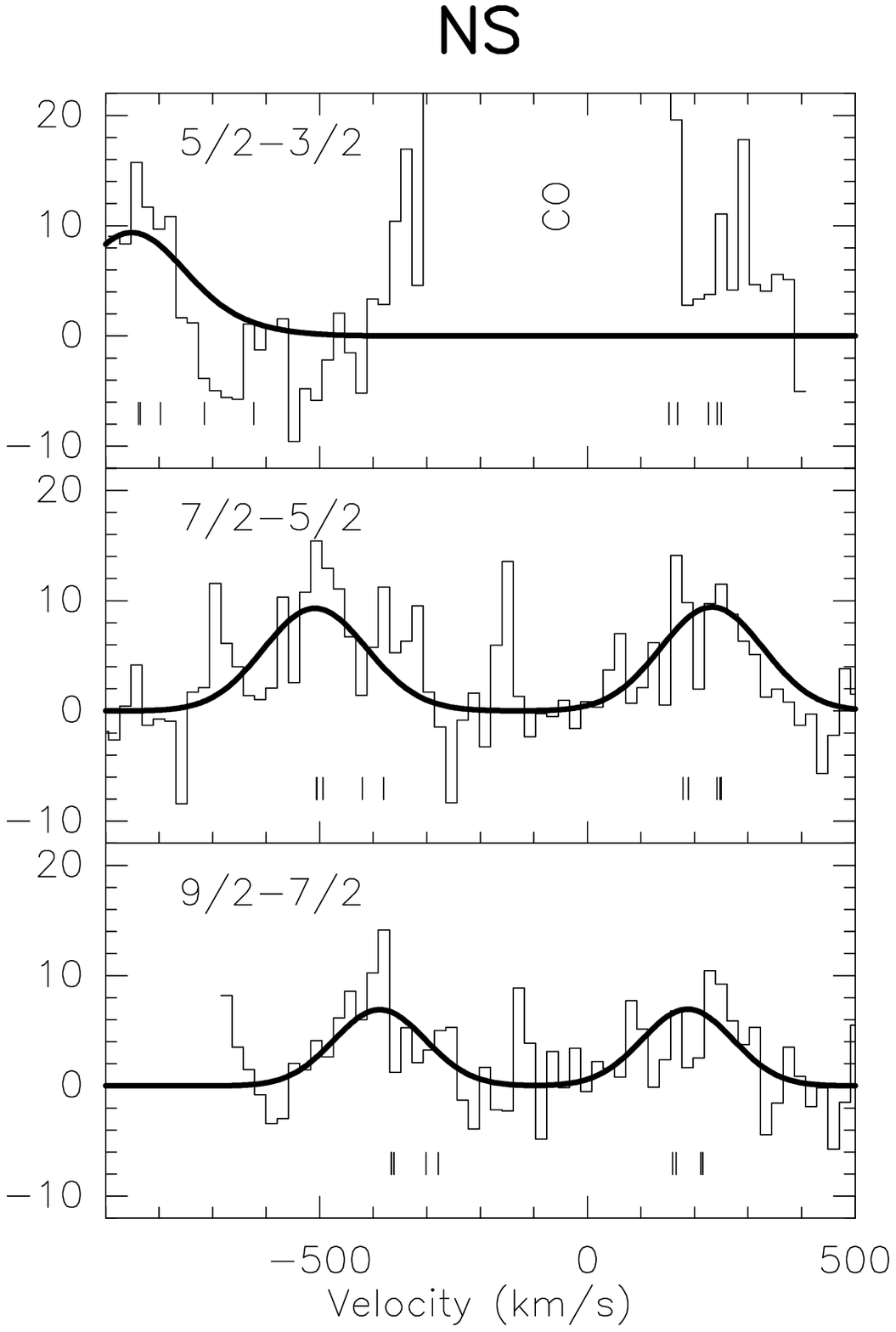}
			\\[5 pt]
			\includegraphics[width=\linewidth]{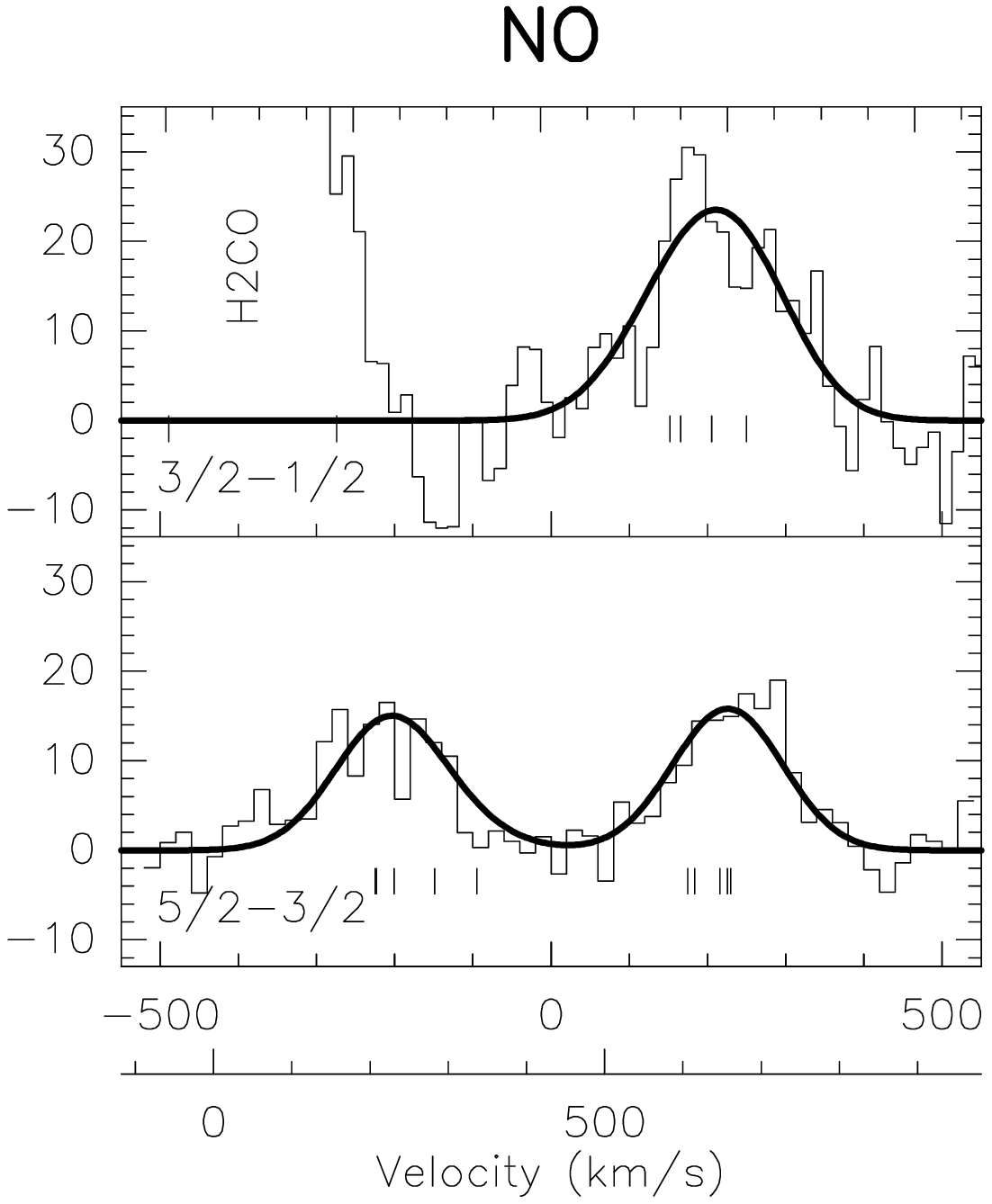}
	\end{minipage}
  \caption {Spectra of the SO$_2$, NS and NO transitions observed
  		towards the nucleus of
		\object{NGC\,253} ($\alpha_{1950}=00^{\rm h}45^{\rm m}06.0^{\rm s} ,\delta_{1950}=-25\degr33\arcmin45\arcsec$).
		Velocity resolutions are 10, 15 and 20 km\,s$^{-1}$ respectively.
		Ticks indicate the expected position of HF components of NS and NO.
		The $T_{\rm MB}$ scale of the y-axis is in mK.}
		\label{fig:transitions}
\end{figure}

As spectrometers we used two 256$\times$4MHz filterbanks for the 2\,mm transitions
and two 512$\times$1MHz filterbanks for the lines at 1.5 and 3 mm. The SIS receivers were tuned to
SSB with an image band rejection $>10$\,dB.
Beam sizes were 21$\arcsec$ (at 3\,mm), 17$\arcsec$ (at 2\,mm) and 10$\arcsec$ or 12$\arcsec$ (at 1.5\,mm).
Spectra were calibrated with the standard dual load system.
Intensities are given on a main-beam brightness temperature scale ($T_{\rm MB}$).
Linear baselines were removed from the spectra and the
resulting profiles are shown in Fig.~\ref{fig:transitions}.

We detected five
transitions of SO$_2$ in the 2\,mm band, three groups of hyperfine components of NO and five of NS in the
2, 3 and 1.5\,mm atmospheric windows.
Profiles are consistent with the frequencies in the spectral line catalogues of Lovas (\cite{Lovas}) and Pickett et al. (\cite{Pickett}).
We checked that no other known molecular transitions, both in the signal and the image band, significantly
contaminate the observed features. The only line partially blended is
SO$_2$ at 135\,GHz that overlaps with $^{34}$SO (see Fig. ~\ref{fig:transitions}).

\section{Analysis}
\subsection{Derived line parameters for SO$_2$, NS and NO}
To derive the SO$_2$ line parameters, we have fitted a single Gaussian to the observed profiles.
Fig.~\ref{fig:transitions} shows the Gaussian fits superposed on the observed spectra, and Table~\ref{tab:fits} presents the line
parameters derived from the fit.

Fitting the NS and NO profiles is more complex as their ground state
presents $\Lambda$-doubling. Thus each rotational level $J$ is split into two single rotational levels with opposite parity.
We use the usual notation to label, in Table~\ref{tab:fits}, the lower and upper series of single rotational levels as $e$ and $f$ for NS, and $\Pi^+$ and $\Pi^-$ for NO.
Rotational levels are further split into hyperfine (HF) components (Gerin et al. \cite{Gerin92}). Fig.~\ref{fig:transitions} indicates 
the location of the HF components, which are unresolved due to the broad intrinsic linewidth of the emission from the nucleus of \object{NGC\,253}.
Single Gaussian fits cannot be used in this case.
We fitted the HF components simultaneously with a comb of Gaussian profiles with identical width and the relative frequency and line
intensities fixed to the spectroscopic parameters of NS and NO.
Table~\ref{tab:fits} shows the derived line parameters.

\begin{table}[]
\caption{Parameters derived from Gaussian fit to SO$_2$ transitions and hyperfine structure fitting for NO and NS transitions.}
\begin{center}
\scriptsize
\begin{tabular}{l l l l l l}
\multicolumn{6}{c}{\large SO$_2$} \\
\\[-6 pt]
\hline
\\[-6 pt]
$\nu$        & Transition  & $\int{T_{\rm MB}{\rm d}v}$   & $V_{\rm LSR}$&  $\Delta v_{1/2}$  & $T_{\rm MB}$ \\
(MHz)        &             &  mK\,km\,s$^{-1}$  & km\,s$^{-1}$ &  km\,s$^{-1}$      & mK         \\
\\[-6 pt]
\hline
\\[-6 pt]
134004.8     &  8(2,6)-8(1,7) & 380  ( 90) & 242             &   91  &  3.9 \\
135696.0     &  5(1,5)-4(0,4) & 1000 (200) & 245$^a$         &  140$^a$&  5.4 \\
140306.1     &  6(2,4)-6(1,5) & 650  ( 80) & 248             &  115  &  5.3 \\
146605.5     &  4(2,2)-4(1,3) & 1080 (230) & 241             &  148  &  6.9 \\
151378.6     &  2(2,0)-2(1,1) & 800  (200) & 225             &  158  &  4.8 \\
\\[-6 pt]
\hline
\multicolumn{6}{c}{} \\
\multicolumn{6}{c}{{\large NS}\,\,$^{b}$} \\ 
\\[-6 pt]
\hline
\\[-6 pt]
$\nu$        & Transition  & $\int{T_{\rm MB}{\rm d}v}$ & $V_{\rm LSR}$&  $\Delta v_{1/2}$  & $T_{\rm MB}$ \\
(MHz)        &  $J-J'$     &  mK\,km\,s$^{-1}$& km\,s$^{-1}$ &  km\,s$^{-1}$      & mK         \\
\\[-6 pt]
\hline
\\[-6 pt]
115556.2   & $\frac{5}{2}-\frac{3}{2}\,f $&     3400 (500) &       220$^a$ &          250$^a$   &          5.6 \\
161297.2   & $\frac{7}{2}-\frac{5}{2}\,e$ &     2500 (700) &       199     &          224       &          4.3 \\
161697.2   & $\frac{7}{2}-\frac{5}{2}\,f$ &     3400 (600) &       249     &          278       &          4.7 \\
207436.2   & $\frac{9}{2}-\frac{7}{2}\,e$ &     1500 (350) &       220$^a$ &          250$^a$   &          2.2 \\
207834.9   & $\frac{9}{2}-\frac{7}{2}\,f$ &     1800 (350) &       220$^a$ &          250$^a$   &          2.6 \\
\\[-6 pt]
\hline
\multicolumn{6}{c}{} \\
\multicolumn{6}{c}{{\large NO}\,\,$^{b}$} \\ 
\\[-6 pt]
\hline
\\[-6 pt]
$\nu$        & Transition  & $\int{T_{\rm MB}{\rm d}v}$ & $V_{\rm LSR}$&  $\Delta v_{1/2}$  & $T_{\rm MB}$ \\
(MHz)        &  $J-J'$     &  mK\,km\,s$^{-1}$& km\,s$^{-1}$ &  km\,s$^{-1}$      & mK         \\
\\[-6 pt]
\hline
\\[-6 pt]
150176.5   & $\frac{3}{2}-\frac{1}{2}\,\Pi^+$    &   4700 (700)      &    244     &   157    &   13.8 \\
250436.8   & $\frac{5}{2}-\frac{3}{2}\,\Pi^+$    &   3100 (320)      &    250     &   150    &    5.4 \\
250796.4   & $\frac{5}{2}-\frac{3}{2}\,\Pi^-$    &   2180 (330)      &    215     &   147    &    6.1 \\
\\[-6 pt]
\hline
\end{tabular}
\end{center}
\label{tab:fits}
$^{a}$ This parameter was fixed in the fit.\\
$^{b}$ transitions and integrated intensity refer to the whole group of HF components while $T_{\rm MB}$ and frequency refer to the main component.\\
\end{table}

\subsection{Column Densities and Rotational Temperatures}

\begin{figure*}[!th]
\begin{center}
\includegraphics[width=0.24\linewidth, angle=-90]{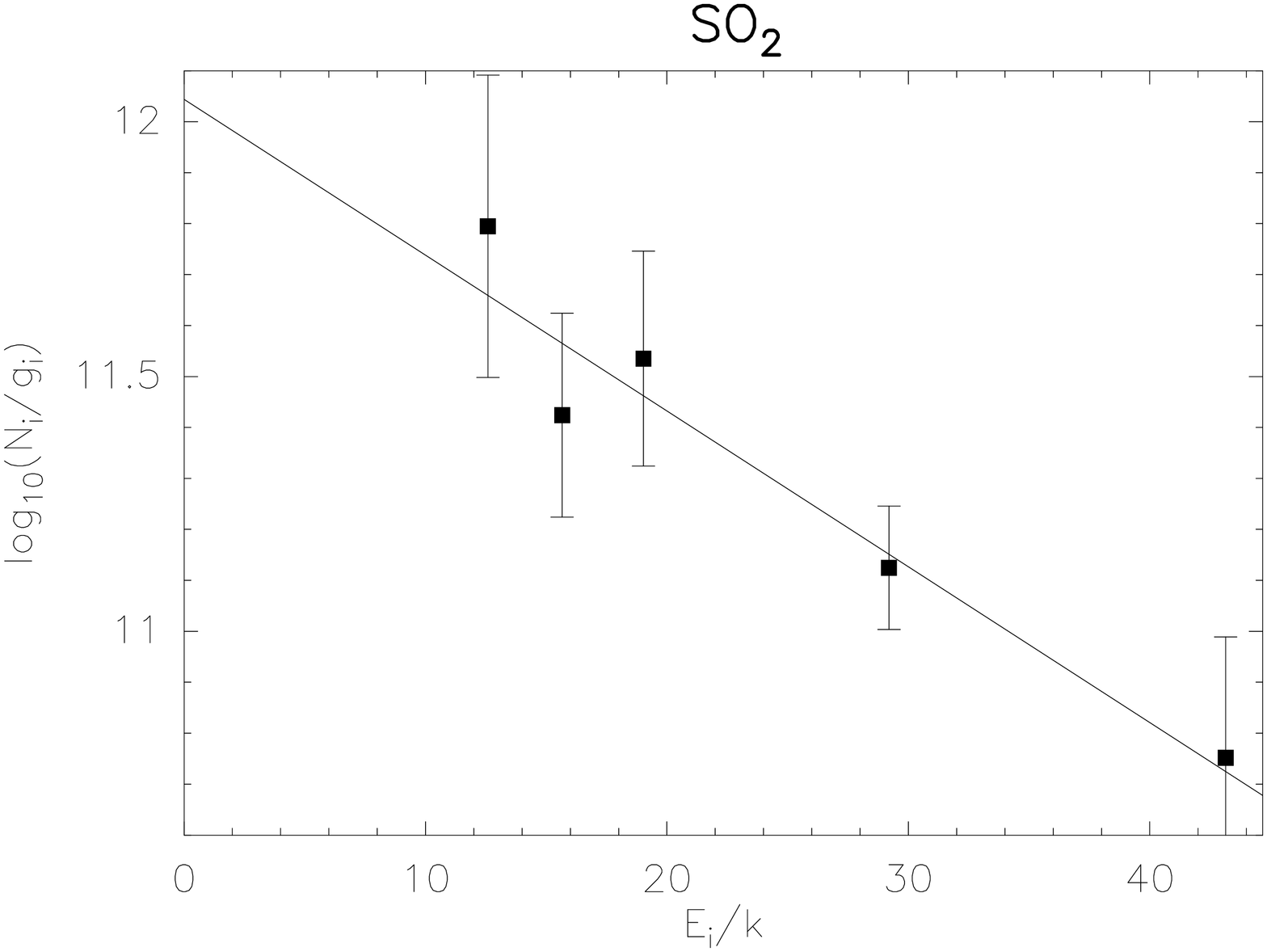}
\includegraphics[width=0.24\linewidth, angle=-90]{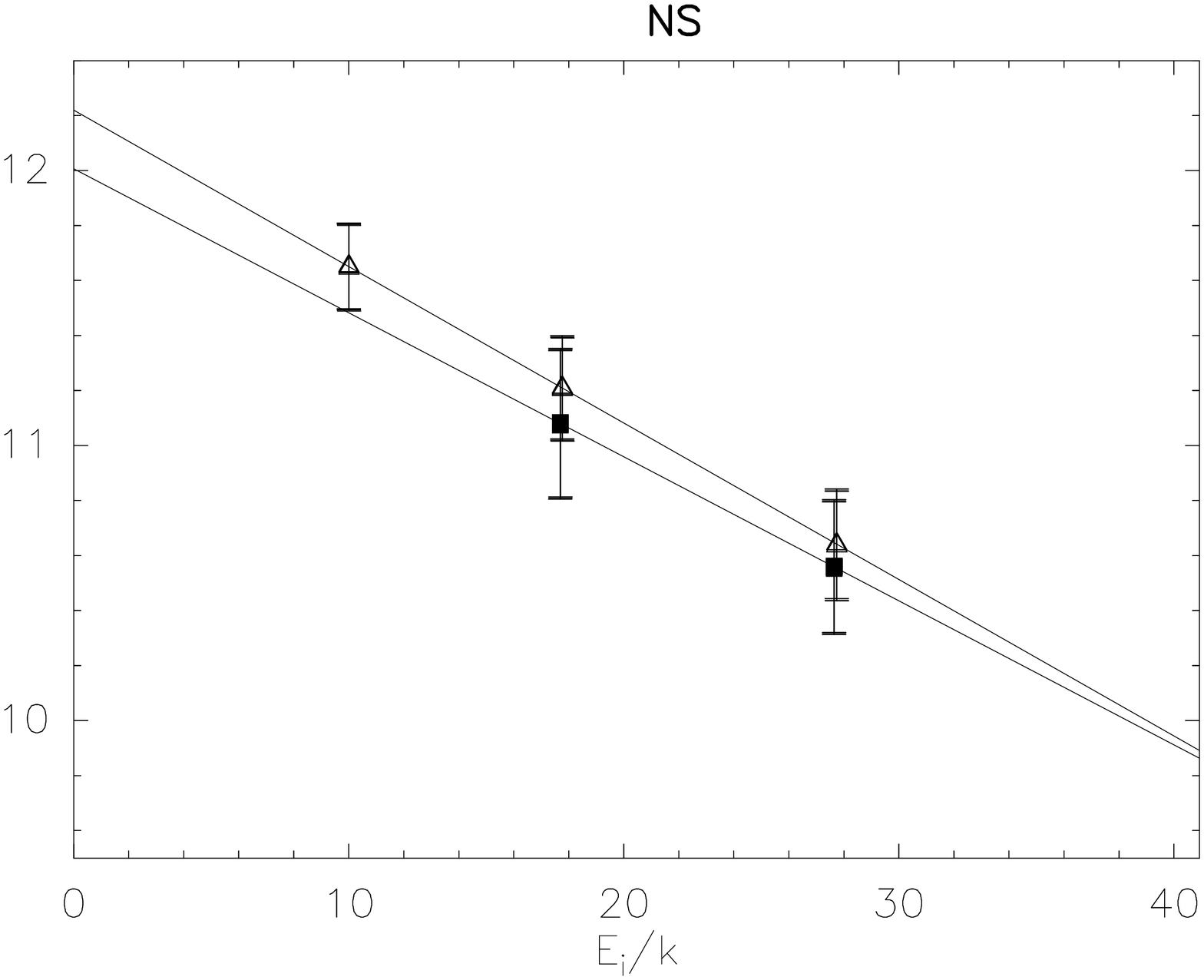}
\includegraphics[width=0.24\linewidth, angle=-90]{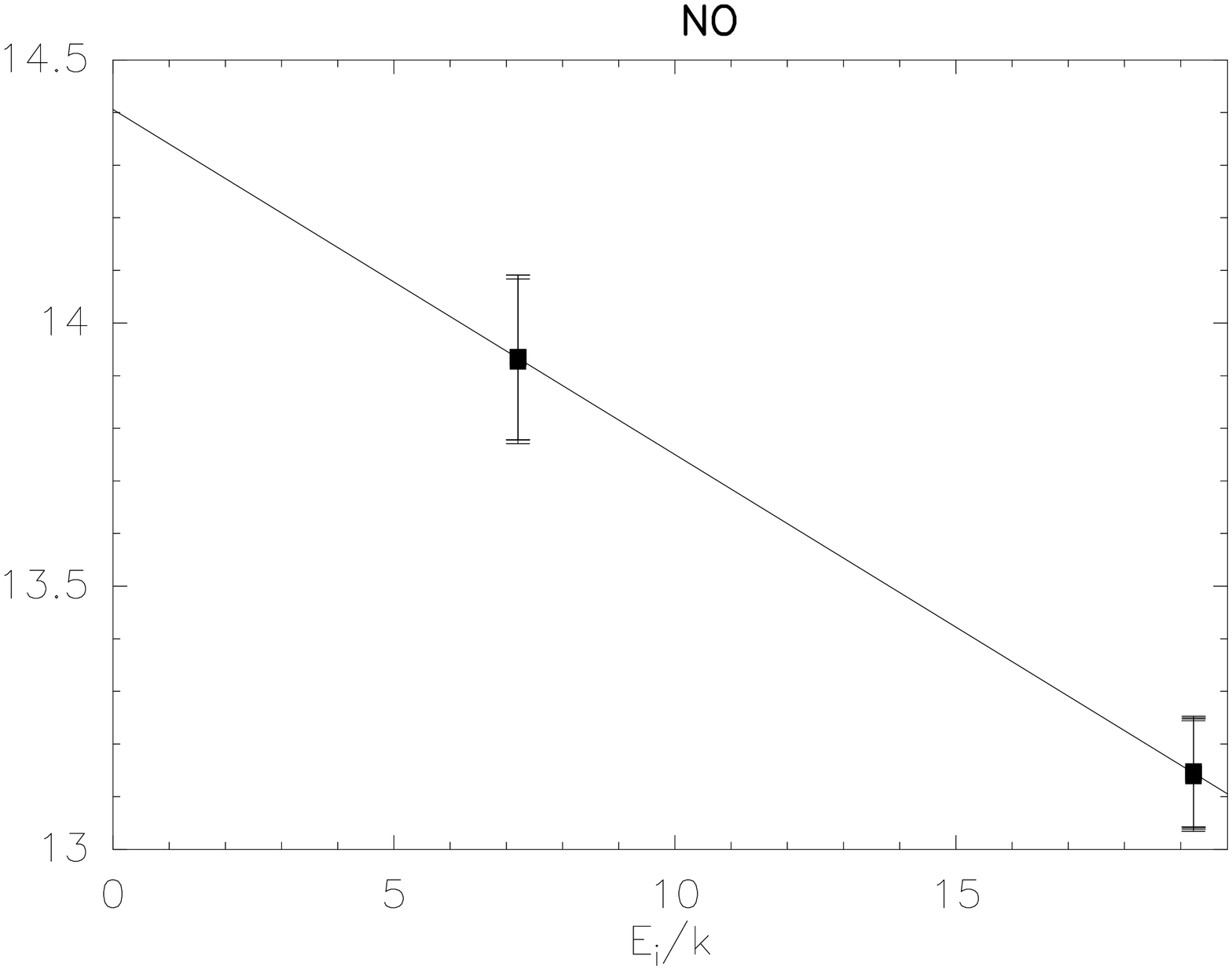}
\caption {Population diagrams for the SO$_2$, NS and NO molecules.
For NS, the $\Pi^+$ (squares) and $\Pi^-$ series (triangles) are plotted and fitted separately.
For NO, only the $e$-series is shown.
}
\label{fig:diagrams}
\end{center}
\end{figure*}

In order to derive reliable physical conditions when comparing line intensities measured with different beam sizes ($\theta_b$), we have to make assumptions on the overall extent ($\theta_s$)
of the nuclear emission of SO$_2$, NO and NS. According to measurements of CO and CN transitions with the different beam sizes of the 30m telescope and
SEST, we derive an extent of the emitting region of $\sim 20\arcsec$ (240\,pc at a distance of 2.5\,Mpc; Mauersberger et al. \cite{Mauers03})
This is in agreement with the size obtained by Mauersberger et al. (\cite{Mauers03}) from
the high angular resolution interferometric maps of CS $J=2-1$ of Peng et al. (\cite{Peng}).
Thus, we can obtain the source averaged brightness temperatures \mbox{($T_{\rm B} = T_{\rm MB}\,\frac{\theta_s^2+\theta_b^2}{\theta_s^2}$)}
over 20$\arcsec$ and derive source averaged column densities in the upper levels of the observed transitions assuming optically thin
emission.
In the case of NS and NO, this was done independently for
each series.
Fig.~\ref{fig:diagrams} shows the population diagrams for each molecule where the level population can be described by a single
rotational temperature represented as a solid straight line.

The derived $T_{\rm rot}$ as well as total column densities and estimated fractional abundances
are summarized in Table~\ref{tab:MolecDensity}.
To determine the abundance, we have assumed the H$_2$ column density of $1.7\,10^{22}\,\rm cm^{-2}$ derived by Mauersberger et al. (\cite{Mauers03})
from their $^{13}$CO measurements. The uncertainty of a factor of 2 in the H$_2$ column density (Mauersberger et al. \cite{Mauers03}) does not critically
affect the conclusions of this work.
In general, the fractional abundances are high with the most outstanding case of NO with an abundance larger than $10^{-7}$.
For the NS molecule, parameters derived for each series are:
$N_{\rm NS}(e)=4.4\,10^{13}\,{\rm cm}^{-2}$, $T_{\rm rot}(e)=8$\,K and $N_{\rm NS}(f)=6.4\,10^{13}\,{\rm cm}^{-2}$, $T_{\rm rot}(f)=7.6$\,K.
No substantial difference is found between series, in agreement with the result of Gerin et al. (\cite{Gerin92}) derived from NO in different
Galactic molecular clouds.

\begin{table}[]
\caption{Source averaged column densities, rotational temperature and abundances.}
\begin{tabular}{l l l l l}
\hline
\\[-10 pt]
Molecule   & $N$            & $T_{\rm rot}$  & [X]/[H$_2$]\,$^a$              \\
           & (cm$^{-2}$)    & (K)            &                                \\
\\[-10 pt]
\hline
\\[-10 pt]
SO$_2$ & $7\,10^{13}$       &  14 (9)      &  $4\,10^{-9}$                  \\
NS     & $5\,10^{13}$       &  8 (1)       &  $3\,10^{-9}$                  \\
NO     & $5\,10^{15} $      &  7 (2)       &  $3\,10^{-7}$                  \\
\\[-10 pt]
\hline
\end{tabular}
\label{tab:MolecDensity}
\\
$^a$ The uncertainty is a factor of 2 in both directions, due to the uncertainty of the assumed $N$(H$_2$)=$1.7\,10^{22}\,\rm cm^{-2}$ (Mauersberger et al. \cite{Mauers03})\\
\end{table}

\begin{table}[]
\caption{Fractional abundances compared with similar studies in molecular clouds.}
\begin{tabular}{l l l l }
\hline
\\[-10 pt]
Source                &            [SO$_2$]/[H$_2$]  &            [NS]/[H$_2$]    &                [NO]/[H$_2$]   \\
\\[-10 pt]
	              &  $10^{-8}$                   & $10^{-9}$                  &   $10^{-8}$                  \\
\hline
\\[-10 pt]
\object{NGC\,253}              & 0.4                          & $3        $                &  30                          \\
\\[-10 pt]
\hline
\\[-10 pt]
\multicolumn{4}{l}{\bf Dark Clouds}\\
\object{L134N}                 & $0.4$$^{a}$                  & $0.2$--$0.6$ $^{b}$        &  20      $^{c}$              \\
\object{TMC1}                  & $<0.1$$^{a}$                 & $0.7$--$1.2$ $^{b}$        &  2.7     $^{c}$              \\
\multicolumn{4}{l}{\bf Hot Cores}\\
\object{Orion Hot Core}        & $9.4$$^{d}$                  &  $0.4$ $^{e}$              &  30      $^{c}$                \\
\object{Sgr\,B2(N)}              & $3$$^{f}$                    & $10$  $^{f}$               &  $20$ $^{f}$                 \\
\object{Sgr\,B2(M)}              & $40$$^{f}$                   & $0.03$ $^{f}$              &  $30$ $^{f}$                 \\
\multicolumn{4}{l}{\bf PDRs}\\
\object{Orion Bar}             & $0.01$$^{g}$                 &                            &  $0.2$ $^{g}$                \\
\hline
\end{tabular}
\label{tab:CloudsCompare}
\\
$^{a}$ Ohishi et al. (\cite{Ohishi});
$^{b}$ McGonagle et al. (\cite{McGonagle94});
$^{c}$ Gerin et al. (\cite{Gerin92}, \cite{Gerin93});
$^{d}$ Charnley (\cite{Charnley});
$^{e}$ McGonagle et al. (\cite{McGonagle97});
$^{f}$ source averaged Nummelin et al. (\cite{Nummelin});
$^g$  Jansen et al. (\cite{Jansen}).\\
\end{table}

\section{Discussion}
To establish the mechanism driving the chemistry in \object{NGC\,253}, we compare in Table~\ref{tab:CloudsCompare} the measured
fractional abundances of SO$_2$, NS and NO with those in
prototypical molecular Galactic clouds dominated by different types of chemistry.
We have selected two dark clouds to illustrate the ion-molecule chemistry associated with quiescent gas, three hot cores to illustrate
grain surface and shock chemistry associated to massive protostars and one photodissociation region (PDR) to illustrate the UV dominated
chemistry produced by OB stars in the main sequence.

Chemistry dominated by PDRs can be ruled out for \object{NGC\,253} since the abundances of SO$_2$ and NO are much larger than those in
Galactic PDRs.
It is remarkable that only one source in Table~\ref{tab:CloudsCompare}, the hot core in \object{Sgr\,B2(N)}, shows larger or similar abundances
than those observed in \object{NGC\,253}.
This is surprising given the expected beam dilution in \object{NGC\,253} for molecules with a high dipole moment like NS and SO$_2$.
However, it is very unlikely that hot cores dominate the emission of the three molecules since the low rotational temperatures derived
are inconsistent with the  high excitation temperature ($>$70\,K) expected from the dense hot cores.
This suggests that the origin of the large abundance of these molecules must be related to a hot core like chemistry, but
in regions with moderate densities in which molecules are subthermally excited.
Molecular clouds in the center of the Galaxy show hot core like chemistry produced by the ejection of molecules from the grains by large
scale shocks (see e.g. Mart\'{\i}n-Pintado et al. \cite{MartinP01}).
Chemistry driven by large scale C shocks has been proposed to explain the high temperatures and large abundances of NH$_3$ and SiO
(Garc\'{\i}a-Burillo et al. \cite{Burillo}, Mauersberger et al. \cite{Mauers03}).
We will now discuss in detail how the abundance of the molecules reported in this paper fits into this picture.

\subsection{\rm SO$_2$}
Towards \object{NGC\,253}, we find a SO$_2$ abundance between that found in dark clouds and hot cores.
The relatively high rotational temperature of SO$_2$ rules out that the bulk of its emission arises from dark clouds with their typical kinetic
temperature of 10\,K.
Within the context of hot core chemistry, SO$_2$ is
rapidly formed from the H$_2$S evaporated from grain-mantles when $T$$<$200\,K. The same type of chemistry is expected if
H$_2$S is injected into the gas phase by shocks.
The observations of SO$_2$ towards \object{Sgr\,B2} by Cummins et al. (\cite{Cummins}) support this idea.
They found that SO$_2$ emission arises from two components with very different rotational temperatures, the first one
associated with the \object{Sgr\,B2} hot core ($T_{\rm rot}$=310\,K) and a second, subthermally excited one, with a much lower temperature of 26\,K
associated with the envelope.
The excitation temperature is somewhat larger in the envelope of \object{Sgr\,B2} than in \object{NGC\,253}. However, the latter is averaged over a much larger region.
We therefore conclude that SO$_2$ arises from the high temperature component observed in NH$_3$ by Mauersberger et al. (\cite{Mauers03}).

\subsection{\rm NO}
The large NO fractional abundance averaged over an almost 200\,pc region suggests that NO emission is fairly widespread through
the whole nucleus of \object{NGC\,253}. If this emission would arise from dark clouds, all the clouds in the nucleus of \object{NGC\,253} must be of the
\object{L134N} type, contrasting with what we find in the solar vicinity where most of the clouds are like \object{TMC1}.
However, the lack of SiO emission in dark clouds (Ziurys et al. \cite{Ziurys}) suggests that large scale shocks
are the origin of the NO abundances observed in \object{NGC\,253}.
Like in the center of our Galaxy, NO emission from \object{NGC\,253} could arise from warm moderately dense gas, such as that of the envelope
of \object{Sgr\,B2} which shows a particular chemistry due to strong shocks and/or irradiation by hard X\,rays (Mart\'{\i}n-Pintado \cite{MartinP}, \cite{MartinP00}).
It is worth noting that models of X\,ray chemistry (Lepp \& Dalgarno \cite{Lepp})
predict an enhancement of the NO abundance ($10^{-6}-10^{-7}$) as observed in \object{Sgr\,B2} and \object{NGC\,253}.
Both \object{NGC\,253} and the envelope of \object{Sgr\,B2}
have enhanced SiO and NO abundances
suggesting a similar origin.

\subsection{\rm NS}
The NS abundance in \object{NGC\,253} is higher than in most Galactic dark cloud cores and hot cores.
McGonagle \& Irvine (\cite{McGonagle97}) studied the excitation of NS in several molecular clouds and found low excitation temperatures and moderate
abundances in the hot and very dense regions as
expected from the large destruction rate of this radical under hot core conditions ($n_{\rm (H_2)}>10^6\,{\rm cm}^{3}, T_{\rm k}>100\,$K).
Chemical models of hot cores indicate an enhancement of the NS abundance relative to CS in the presence of C shocks (Viti et al. \cite{Viti}).
The high NS/CS ratio of ~0.4 (with $X$(CS)=4\,10$^{-9}$; Mauersberger et al. \cite{Mauers03}) agrees with the prediction of these models in
which a shock passed $10-20\,10^{3}$ years after the onset of radiative heating in the hot core.
This favors large scale shocks as the most likely explanation for the NS abundance in \object{NGC\,253}.
It is interesting
to note that this NS/CS ratio is an order of magnitude above the ratio measured by Hatchell et al. (\cite{Hatchell}) towards six hot cores.
\\
\\
The origin of the large scale shocks is not yet clear. High angular resolution images of nuclear SiO
emission in \object{NGC\,253} suggest different origins (Garc\'{\i}a-Burillo et al. \cite{Burillo}).
In the inner circumnuclear disk, shocks from a dense molecular outflow driven by
massive protostar dominates.
In view of the low excitation temperatures derived for the new molecules, the
bulk of the emission is probably not related to this origin.
The other proposed causes, shocks in the outer Lindblad resonance as a consequence of a barred potential or
generated by mass ejection from the disk, could explain the abundances and excitation temperatures.

Interferometric observations of SO$_2$, NO and NS are needed to measure their spatial distribution and to establish the origin of the
large abundances of these molecules in the nuclear region of NGC 253.

\begin{acknowledgements}
J.\,M.-P. has been partially supported by the Ministerio de Ciencia Y Tecnolog\'{\i}a with grant ESP2002-01627 and AYA2002-10113E.
\end{acknowledgements}

\newpage

\end{document}